\documentclass{aa}

\usepackage{natbib}
\usepackage{epsfig}
\usepackage{epsf}
\usepackage{color}
\definecolor{red}{rgb}{0.7,0,0}
\definecolor{blue}{rgb}{0,0,0.7}
\def\correc#1{{#1}}

\def\ergcms{erg~cm$^{-2}$~s$^{-1}$ }
\def\nh{N$_\mathrm{H}$}
\def\cm2{cm$^{-2}$}

\def\integral{{\it{INTEGRAL}}}
\def\swift{{\it{Swift}}}
\def\chisq{$\chi^2_\nu$}

\usepackage{longtable}
\usepackage{natbib}
 \usepackage[figuresright]{rotating}
\usepackage{lscape}
\usepackage{txfonts}
\begin{document}
   \title{\swift\ follow-up observations of 17 INTEGRAL sources of uncertain or unknown nature}

   \author{J. Rodriguez
          \inst{1}
          \and
          J.A. Tomsick \inst{2}
	  \and
          S. Chaty\inst{1}}

   \offprints{J. Rodriguez}
\authorrunning{Rodriguez, Tomsick, Chaty}
   \institute{Laboratoire AIM, CEA/DSM - CNRS - Universit\'e Paris Diderot, IRFU/SAp,
 Centre de Saclay, F-91191 Gif-sur-Yvette, France\\
              \email{jrodriguez@cea.fr}
         \and
             Space Sciences Laboratory, 7 Gauss Way,
University of California, Berkeley, CA 94720-7450, USA\\
             }

   \date{}

 
  \abstract
   {The positional accuracy of the IBIS telescope on-board \integral, albeit unprecedented
 in the $>20$~keV range, is still not good enough to identify many hard X-ray sources 
discovered by \integral. This indeed prevents counterparts at other wavelengths 
 from being found, which is the unique way to unveil the true nature of these sources.}
   {We continue the work of trying to reveal the nature of these hard X-ray sources. This 
is done by analysing X-ray data collected with focusing X-ray telescopes, with the primary 
goal of discovering soft X-ray counterparts of the \integral\ sources to provide an accurate 
X-ray position.  With few arcsec accuracy, we can identify counterparts at infrared and optical 
wavelengths.}
{We analysed data from observations of 17 \integral\ sources made with the \swift\ satellite.
The X-ray images obtained by the X-Ray Telescope instrument allowed us to refine the position 
of the hard X-ray sources to an accuracy of a few arcsec. We then browsed the online catalogs 
(e.g., NED, SIMBAD, 2MASS, 2MASX, USNO) to search for counterparts at other wavelengths. We 
also made use of the X-ray spectral parameters to further distinguish between the various
possibilities.}
{For 13 sources, we find the X-ray counterpart without any ambiguity. For these, we provide 
the position with arcsec accuracy, identify possible infrared and optical counterparts 
(when found), give the magnitudes in those bands and in the optical and UV as seen with the 
\swift\/UVOT telescope when observations are available.  We confirm the previously suggested 
associations and source types for IGR~J03532$-$6829, J05346$-$5759, J10101$-$5654, J13000+2529, 
J13020$-$6359, J15479$-$4529, J18214$-$1318, and J23206+6431. We identify IGR~J09025$-$6814 
as an AGN for the first time, and we suggest that it may be a Seyfert 2.  We suggest that 
IGR~J05319$-$6601, J16287$-$5021, J17353$-$3539 and J17476$-$2253 are X-ray binaries, with
J05319$-$6601 being located in the LMC and the other three possibly being HMXBs in our 
Galaxy.  For IGR~J15161$-$3827 and J20286+2544, we find several possible X-ray counterparts 
in the IBIS error region, and we discuss which, if any, are the likely counterparts.  Both 
are likely AGNs, although the latter could be a blend of two AGNs. For IGR~J03184$-$0014 and 
J19267+1325, we find X-ray sources slightly outside the IBIS error circle.  In the former, we 
do not favour an association of the \swift\ and \integral\ source, while it is very likely that  
IGR~J19267+1325 and the \swift\ source are the same.}
   {}


   \maketitle
%

\section{Introduction}
Since its launch, the INTErnational Gamma-Ray Astrophysics Laboratory (\integral) has 
detected about 500 sources as reported in a recent version of its source catalog 
\citep{bird07,bodaghee07}.  A large number of the sources were either not well-studied
or had not been detected prior to \integral.  In this paper, we will refer to them as 
`IGRs'\footnote{An up-to-date online catalog of all IGRs can be found at 
http://isdc.unige.ch/$\sim$rodrigue/html/igrsources.html}.  Although $\sim$arcmin accuracy 
is achieved for source positions with IBIS/ISGRI \citep{lebrun03}, a level which is 
unprecedented in the $>20$~keV range, this is not sufficient to unveil counterparts 
at other wavelengths (optical, infrared (IR) and radio), which is the best way to 
reveal the true nature of the IGRs. \\
\indent In a recent paper, \citet{bodaghee07} collected known parameters (e.g., the 
absorption column density, \nh, the pulse period for Galactic sources with X-ray 
pulsations, the redshift for AGN, etc.) of all sources detected by \integral\ during 
the first four years of activity.  Their catalog, 
however, contains a large number of IGRs whose high energy position is accurate 
at just the arcmin level, which therefore prevents their true nature from being
known.  In some cases, a tentative identification is given, mainly when an AGN is 
found within the \integral/ISGRI error circle, but this is far from being secure as 
other possible counterparts usually lie in the few arcmin ISGRI error regions.\\
\begin{table}[htbp]
\caption{Journal of the \swift\ observations analysed in this paper.}
\begin{tabular}{lllll}
\hline
\hline
Source Id & Id  & Date Obs & Tstart & Exposure\\
(IGR)     &      &          &  (UTC) &  (s) \\
\hline
J03184$-$0014  &  00030995001  &   2007-11-07 & 00:12:58   &  9192 \\
J03532$-$6829  &  00037303001  &   2008-07-02 & 13:59:56   &  2405 \\
J05319$-$6601  &  00036094001   &   2007-01-07 & 07:16:32   &  1395 \\
               & 00036094002   &   2008-01-01 & 00:05:08   &  17649 \\
J05346$-$5759  &  00037120001   &   2007-11-13 & 01:29:04   &  5926 \\
               &  00037120002   &   2007-12-25 & 12:08:50   &  2762 \\
               &  00037120003   &   2007-12-31 & 15:41:50   &  6966 \\
J09025$-$6814  &  00037312001   &   2008-02-07 & 20:00:42   &  1054 \\
               &  00037312002   &   2008-03-02 & 00:46:22   &  4119 \\
               &   00037312003   &   2008-03-18 & 02:23:28   &  2529 \\
               &  00037312004   &   2008-05-08 & 07:25:07   &  2269 \\
J10101$-$5654  &  00030356001   &   2006-01-12 & 08:07:43   &  1201 \\
J13000+2529    &  00036818001   &   2008-02-23 & 09:56:41   &  558 \\
               &  00036818002   &   2008-02-22 & 06:43:11   &  744 \\
J13020$-$6359  &  00030966001   &   2007-07-07 & 14:35:41   &  2705 \\
               &  00030966002   &   2007-07-09 & 13:27:01   &  5126 \\
               &  00030966003   &   2007-07-11 & 07:09:27   &  5512 \\
               &  00030966004   &   2007-07-13 & 16:49:45   &  5951 \\
J15161$-$3827  &  00036663001   &   2008-01-25 & 23:38:01   &  7808 \\
               &  00036663002   &   2008-01-27 & 01:21:41   &  5309 \\
J15479$-$4529  &  00037149001   &   2007-06-23 & 14:49:57   &  346 \\
               &  00037149002   &   2007-06-24 & 00:28:26   &  3968 \\
               &  00037149003   &   2007-06-26 & 00:41:28   &  983 \\
               &  00037149004   &   2008-01-25 & 01:01:51   &  4758 \\
               &  00037149005   &   2008-06-25 & 01:19:05   &  2580 \\
	       &  00037149006   &   2008-06-26 & 07:50:53   &  1685 \\
J16287$-$5021  &  00037074001   &   2008-07-11 & 17:20:34   &  1944 \\
J17353$-$3539  &  00311603004   &   2008-05-28 & 00:38:42   &  4540 \\
               &  00311603005   &   2008-06-04 & 23:56:39   &  184 \\
	       &  00311603006   &   2008-06-05 & 06:14:18   &  4368 \\
               &  00311603008   &   2008-06-14 & 03:48:37   &  3869 \\
               &  00311603009   &   2008-07-12 & 04:49:18   &  8713 \\
J17476$-$2253  &  00036656001   &   2008-07-03 & 20:16:28   &  1142 \\
J18214$-$1318  &  00035354001   &   2006-02-11 & 15:30:34   &  6285 \\
J19267+1325    &  00037062001   &   2007-07-20 & 11:15:50   &  4312 \\
J20286+2544    &  00030722001   &   2006-06-03 & 14:44:55   &  6876 \\
               &  00035276001   &   2005-12-16 & 01:19:43   &  4525 \\
               &  00035276002   &   2006-03-23 & 00:23:43   &  4597 \\
               &  00035276003   &   2006-03-28 & 01:20:05   &  921 \\
J23206+6431    &  00031026001   &   2007-11-24 & 00:05:08   &  3978 \\
\hline
\hline
\end{tabular}
\label{tab:log}
\end{table}
\indent In this paper, we continue our work of identifying the unknown IGRs
that we started soon after the discovery of the first IGRs.  A first step
is to provide an $\sim$arcsec position with soft X-ray telescopes such as 
{\it {XMM-Newton}}, {\it{Chandra}} 
\citep[e.g.,][]{rodriguez03, rodriguez06, tomsick06, tomsick08}, and also 
\swift\ \citep[][hereafter paper 1]{rodriguez08}. We then search for 
counterparts at a position consistent with the refined X-ray position of 
a given source. Note that in the case of HMXBs, we also have follow-up 
programmes from ground-based facilities that permit us to further understand 
the nature of a large number of systems \citep{chaty08, rahoui08}. In 
paper 1, we focused on sources that were easily detected with \swift/XRT 
\citep{gehrels04, burrows05}, i.e., sources that were bright enough to be 
detected during single pointings lasting a few ks. In this paper, we report 
on the analysis of \swift\ observations (XRT imaging and spectral analysis 
and UVOT imaging) of seventeen IGRs that either lacked precise arcsec X-ray positions or 
whose Chandra refined X-ray position was very recently published  by 
us \citep{tomsick08,tomsick08b}. We 
also present the identification of IR and optical counterparts obtained from 
online catalogs such as SIMBAD, the United States Naval Observatory (USNO), 
the 2 Micron All Sky Survey point source and extended source 
catalogs\footnote{http://www.ipac.caltech.edu/2mass/} (2MASS and 2MASX, 
\citet{skrutskie06}), and the NASA/IPAC Extragalactic Database 
(NED\footnote{http://nedwww.ipac.caltech.edu/index.html}). 
\correc{It should be noted that although the presence of a bright \swift\ 
source within a given \integral\ error circle renders very likely the 
association between the two sources, there is a non-null probability that 
the two sources are not associated. This is, in particular, exemplified by 
the few cases where several \swift\ sources are found within the \integral\ 
error circle.  Note that this remark is also true for the association between 
the \swift\ sources and the proposed counterpart at other wavelengths. We cannot 
give a general statement about this issue, that would hold for all cases, as there 
is a large range of association probabilities from possible associations to 
nearly certain associations. For all sources, we discuss the likelihood of association
between the \integral, \swift, and counterparts at other wavelengths. 
Dubious cases (as, e.g., multiple possible counterparts) are discussed in more detail. } \\
\indent We start by introducing the \swift\ observations and \correc{briefly 
presenting the} data reduction techniques in Sec.~2.  Then, in Sec.~3, we 
describe the results for each source (position, counterparts, and spectral 
properties) and discuss their possible nature.  We conclude the paper by 
summarising the results in Sec.~4.

\section{Observations and data reduction}
Among all the \swift\ pointed observations of IGRs, we mainly restricted our 
analysis to sources whose fine position and/or \swift\ observations were not 
published anywhere else\footnote{with the exceptions of IGR~J10101$-$5654, 
J18214$-$1318, J16287-5021, and J19267+1325 whose {\it Chandra} positions 
have very recently been published by \citet{tomsick08,tomsick08b}}. 
We used only the 
pointings during which the XRT instrument was in photon counting 
mode since this is
the only mode that provides a fine position.  We also included in our study 
sources for which a {\it{possible}} identification had been given, e.g., 
based on the presence of an AGN in the IBIS error region in existing 
catalogs \citep[see, e.g.,][]{bodaghee07}.  The observing log for our sample
of seventeen sources is reported in Table~\ref{tab:log}. \\
\indent \correc{We reduced the \swift\ data with the {\tt{HEASoft V6.5}} 
software package and the calibration files issued on 2008 May 1 and 2008 June
25 for the UVOT and XRT instruments, respectively.  The reduction steps are 
identical to those presented in paper 1, and follow the standard steps 
described in the XRT users 
guide and UVOT software 
guides\footnote{both available at http://heasarc.gsfc.nasa.gov/docs/swift/analysis/}. 
More specifically, we ran the {\tt{xrtpipeline}} tool with standard screening 
criteria to produce level 2 (i.e., cleaned) event files from the level 1 data 
products. The positions of the sources were obtained with {\tt{xrtcentroid}}. 
We co-added all individual pointings of a given source with {\tt{xselect}}, 
before estimating the source position from the resulting mosaic. We extracted
spectra and light curves with {\tt{xselect}} from a circular region with a radius 
of 20 pixels centred on the best position, while we obtained the background 
products from a source-free circular region with a radius of 40 pixels (see 
also paper 1).  Due to the presence of columns of dead pixels in the XRT, we 
produced ``true'' exposure maps to further correct the ancillary response 
files (see also paper 1). We rebinned the spectra to have at least 20 counts 
per channel which allows for $\chi^{2}$-minimization in the fitting with
{\tt{XSPEC 11.3.2ag}}. When this criterion was not achievable, the Cash 
statistic (hereafter C-statistic, \citet{cash76}) was used instead.}\\
\indent\correc{When available, we analysed the UVOT level 2 data obtained from 
the \swift\ data archive.  We first corrected the aspect for each individual 
UVOT exposure with the {\tt{uvotskycorr}} tool, calculating the aspect correction 
via comparison to the USNO-B1.0
catalogue\footnote{http://tdc-www.harvard.edu/software/catalogs/ub1.html}\citep{monet03}.
Then, we summed the aspect-corrected individual exposures with {\tt{uvotimsum}}, 
and performed the UVOT photometry and astrometry with the {\tt{uvotdetect}} tool.} 

\section{Results}
The refined X-ray positions of the sources detected by \swift\ are reported in 
Table~\ref{tab:position}.  For each source, we searched the 2MASS, 2MASX 
\correc{and the USNO-B1.0} online catalogs for the presence of infrared 
\correc{and/or optical} counterparts within the \swift/XRT error circle. 
Infrared counterparts that are newly identified from this search are reported 
in Table~\ref{tab:ircounterparts}. \correc{The typical positional accuracy for 
the 2MASS sources is 0.5\arcsec\ \citep{skrutskie06}, while that of the USNO-B1.0 
sources is typically 0.2\arcsec\ \citep{monet03}.} \correc{The magnitudes and UV 
positions of the optical and UV counterparts are reported in 
Table~\ref{tab:uvcounterparts}.}\\
\indent \correc{We fitted the source spectra 
with a simple model of an absorbed power law. This provided an acceptable 
representation of the spectra in the large majority of the cases.} The spectral parameters we 
obtained are reported in Table~\ref{tab:spectral}.  The errors on the X-ray spectral 
parameters (including upper limits) are at the 90$\%$ confidence level.
 We discuss in the following subsections the results obtained for each of the 
sources, \correc{including the few cases where a simple absorbed power law is not sufficient, 
or not appropriate to represent the spectra well. To estimate the luminosity of the 
candidate AGN we used H$_0$=75~km/s/Mpc to convert the redshift (of the suggested counterpart)
to distance.} 
\correc{The lower limits on the UVOT magnitudes are given at the $3\sigma$ 
level. The UVOT positional uncertainties are dominated by a 0.5\arcsec\ 
systematic uncertainty (90\% confidence) for each source.}  All X-ray fluxes 
and luminosities are corrected for absorption.  \correc{The absorption due to 
intervening material along the line of sight is first obtained with the 
{\tt{nh}} tool based on the measurements of H~I made by \citet{dickey90}. It 
is also compared to the values obtained from the Leiden/Argentine/Bonn (LAB)  
surveys of Galactic H~I in the Galaxy. The LAB Survey is the most sensitive 
Milky Way H~I survey to date, with the most extensive coverage both spatially 
and kinematically and an angular resolution of 0.6 degrees \citep{kaberla05}. 
For each source, the two values are reported in Table~\ref{tab:spectral} for 
comparison}.  
\begin{table*}[htbp]
\caption{X-ray position (equatorial and Galactic) of the X-ray counterparts to the 17 sources studied with \swift/XRT.}
\label{tab:position}
\begin{tabular}{lllcll} 
\hline\hline             
Name & RA & DEC & Error & l & b \\
(IGR)  & (J2000) & (J2000) & (\arcsec) & (\degr) & (\degr) \\ 
\hline
J03184$-$0014$^\dagger$ & 03h 18m 17.6s & $-$00\degr 17\arcmin 48.1\arcsec & 5.7 & 181.8112 & $-$45.7082 \\
J03532$-$6829 & 03h 52m 57.4s & $-$68\degr 31\arcmin 18.0\arcsec & 3.5 & 282.8102 & $-$40.7968 \\
J05319$-$6601$^\dagger$ & 05h 31m 52.6s & $-$65\degr 59\arcmin 40.2\arcsec & 4.7 & 275.9037 & $-$32.6650  \\
J05346$-$5759& 05h 34m 50.5s & $-$58\degr 01\arcmin 39.3\arcsec & 3.5 & 266.4230 & $-$32.7788   \\
J09025$-$6814$^\dagger$ & 09h 02m 39.4 & $-$68\degr 13\arcmin 38.7\arcsec & 4.8 &  284.1738 & $-$14.1567 \\
J10101$-$5654$^\star$ & 10h 10m 11.9s & $-$56\degr 55\arcmin 31.6\arcsec & 4.3 & 282.2567 & $-$0.6719  \\
J13000+2529$^\dagger$ & 12h 59m 55.0s & +25\degr 28\arcmin 08.8\arcsec & 6.9 & 352.2816 & +87.4774 \\
J13020$-$6359 & 13h 01m 59.2s & $-$63\degr 58\arcmin 06.0\arcsec & 3.5 & 304.0891 & $-$1.1202 \\
J15161$-$3827$^\ddagger$ \#1 & 15h 15m 59.3s & $-$38\degr 25\arcmin 48.3\arcsec & 4.3 & 331.6935 & +16.2381 \\
               \#2    & 15h 16m 29.6s & $-$38\degr 26\arcmin 56.5\arcsec & 4.6 & 331.7689 & +16.1681\\
 \#3$^\dagger$      & 15h 16m 12.7s & $-$38\degr 31\arcmin 02.4\arcsec & 4.7 & 331.6819 & +16.1411  \\
 \#4$^\dagger$      & 15h 15m 45.8s & $-$38\degr 27\arcmin 36.2\arcsec & 4.7 & 331.6380 & +16.2370 \\
J15479$-$4529 & 15h 48m 14.7s & $-$45\degr 28\arcmin 40.4\arcsec & 3.5 & 332.4403 & +7.0228 \\
J16287$-$5021$^\diamond$ & 16h 28m 27.2s & $-$50\degr 22\arcmin 38.3\arcsec & 4.4 & 334.1093 & $-$1.1261 \\
J17353$-$3539 & 17h 35m 23.5s & $-$35\degr 40\arcmin 13.8\arcsec & 3.5 & 353.1445 & $-$1.7401 \\
J17476$-$2253 & 17h 47m 30.0s & $-$22\degr 52\arcmin 43.2\arcsec & 4.8 & 5.3999 & +2.7813 \\
J18214$-$1318$^\star$ & 18h 21m 19.7s & $-$13\degr 18\arcmin 38.2\arcsec & 3.5 & 17.6813 & +0.4856  \\
J19267+1325$^\diamond$ & 19h 26m 27.0s & +13\degr 22\arcmin 03.4\arcsec & 3.7 & 48.8032 & $-$1.5059 \\
J20286+2544$^\ddagger$ \#1 & 20h 28m 34.9s & +25\degr 43\arcmin 59.7\arcsec & 3.9 & 67.0045 & $-$7.5713 \\
            \#2    & 20h 28m 28.7s & +25\degr 43\arcmin 22.5\arcsec & 4.4 & 66.9825 & $-$7.5582 \\
J23206+6431& 23h 20m 36.8s & +64\degr 30\arcmin 42.8\arcsec & 3.8 & 113.3539 & +3.3424  \\
\hline
\hline
\end{tabular}
\begin{list}{}{}
\item[$^\dagger$]Source is very faint, just a very slight excess (very few photons) within IBIS error. 
\item[$^{\star}$]Consistent with the {\it Chandra} position published by \citet{tomsick08}
\item[$^\ddagger$]Several sources within IBIS error
\item[$^\diamond$]Consistent with the {\it Chandra} position published by \citet{tomsick08b}
\end{list}
\end{table*}

\begin{table*}[htbp]
\caption{List of newly identified infrared counterparts in the 2MASS and 2MASX catalogs.}\label{tab:ircounterparts}
\begin{tabular}{llcccc} 
\hline\hline             
Name & Counterpart &\multicolumn{3}{c}{Magnitudes} & Offset from the\\
(IGR)  &  & J & H & K$_\mathrm{s}$  & XRT position (\arcsec)\\ 
\hline
J03184$-$0014 & 2MASS J03181753$-$0017502 &                 &                & 15.2$\pm$0.1 & 2.4\\
J03532$-$6829 & 2MASX J03525755$-$6831167 &  13.22$\pm$0.04 & 12.50$\pm$0.05 & 12.07$\pm0.08$ & 1.5  \\
J05346$-$5759 & 2MASS J05345057$-$5801406 &  14.77$\pm$0.04 & 14.34$\pm$0.05 & 14.11$\pm0.06$ & 1.4  \\
J09025$-$6814 & 2MASX J09023946$-$6813365 &  10.24$\pm$0.01 & 9.50$\pm0.01$  & 9.19$\pm$0.02  & 2.1  \\
J13000+2529   & 2MASS J12595533+2528101   &  10.39$\pm$0.02 & 9.80$\pm0.03$  & 9.68$\pm$0.02  & 4.7  \\
J15161$-$3827 \#1 & 2MASX J15155970$-$3825468 & 12.55$\pm$0.03 & 11.83$\pm$0.03 & 11.34$\pm$0.06 & 4.9  \\
              \#3 & 2MASS J15161246$-$3831041 & 10.45$\pm$0.02 & 10.21$\pm$0.02 & 10.13$\pm$0.02 & 3.5  \\
J15479$-$4529 & 2MASS J15481459$-$4528399 & 13.22$\pm0.03$ & 12.75$\pm$0.03 &  12.53$\pm0.03$ & 1.2 \\
J17353$-$3539 & 2MASS J17352361$-$3540128 & 10.23$\pm0.02$ & 9.03$\pm0.02$ & 8.63$\pm0.03$  & 1.6 \\
J17476$-$2253 & 2MASS J17472972$-$2252448 &                &               & 13.00$\pm0.07$ & 4.2\\ 
J20286+2544 \#1 & 2MASX J20283506+2544001 & 11.31$\pm0.02$ & 10.39$\pm0.02$  & 9.93$\pm0.03$ & 2.3 \\
            \#2 & 2MASX J20282884+2543241 & 10.05$\pm0.01$ & 9.23$\pm0.01$  & 8.87$\pm0.01$  & 2.6 \\
\hline
\hline
\end{tabular}
\end{table*}

\subsection{Confirmations of previously suggested associations}

\paragraph{\bf \object{IGR~J03532$-$6829: }\\}
\citet{masetti06a} suggested an association of the IGR source with PKS~0352$-$686, 
a blazar of BL Lac type at $z$=0.087, based on its location inside the IBIS error circle 
\citep{gotz06} as well as the fact that these objects are known to be strong
emitters of X- and gamma-rays.  The source detected by \swift/XRT is 1.14\arcsec\ 
from the position of \object{PKS~0352$-$686} reported in NED, \correc{further 
strengthening the classification of the IGR source as a BL Lac.}  The extended 
2MASX source that lies within the XRT error circle (Table~\ref{tab:ircounterparts}) 
has already been associated with the BL Lac.  \correc{There is also one USNO-B1.0 
source and a single UVOT source within the \swift\ error circle 
(Table~\ref{tab:uvcounterparts}). The USNO-B1.0 and UVOT sources are at positions 
consistent with the BL Lac object given the $\sim30$\arcsec\ extension of the 2MASX 
source}.  The \swift\ source is coincident with \object{1RXS~035257.7$-$683120} 
which is classified as being a cluster of galaxies in SIMBAD. \\
\indent An absorbed power-law represents the \swift/XRT spectrum well with 
\chisq=0.98 for 63 degrees of freedom (dof). The value of the absorption 
(Table~\ref{tab:spectral}) is compatible with the value of Galactic absorption 
along the line of sight. \correc{This indicates that the source 
is not significantly locally absorbed. This further argues in favour of the hard X-ray source 
being the blazar as these objects do not usually show significant intrinsic absorption.} 
At $z$=0.087, the 2--10 keV luminosity of the source is $\sim2.5\times 10^{44}$~erg/s. 
We note that the extrapolated 20--40 keV flux of the \swift\ spectrum is about 
twice as high as the \integral\ flux \correc{of 0.6 mCrab} reported in \citet{gotz06}.
If the extrapolation of the power-law is valid, then this indicates variability, as
expected in a BL Lac.

\paragraph{\bf \object{IGR~J05346$-$5759: }\\}
Based on positional coincidence and the good agreement between the \integral\ and
{\it ROSAT} spectral shape, \citet{gotz06} suggested that IGR~J05346$-$5759 is the 
hard X-ray counterpart to TW Pic, a Cataclysmic Variable (CV).  There is a unique and 
quite bright XRT source within the IBIS error circle. \correc{TW Pic is the only 
source given in SIMBAD that is within the XRT error circle, where it is also 
associated with the 2MASS source listed in Table~\ref{tab:ircounterparts}.  The 
single source that is found in the USNO-B1.0 catalogue is positionally coincident
with the single detected UVOT source (see Table \ref{tab:uvcounterparts}), indicating
that they are the same source.  We note that the UVOT magnitudes were obtained from 
pointing \#2 for the UVW1 filter and pointings \#1 and \#3 for the other two
filters. The values obtained in the latter two are compatible (within the 0.2 mag errors)
and we report the mean of the two in Table \ref{tab:uvcounterparts}.  These spatial 
coincidences strengthen the association of the XRT source with the CV.} \correc{The fact 
that CVs are known X-ray emitters, and that an increasing number have been seen at X-ray 
energies $>20$ keV, makes the suggested associations between IGR~J05346$-$5759 and TW Pic 
very likely and secure.} \\
\indent We first checked the XRT count rates for variability between the different 
pointings. The source shows some variability between high flux states (up to $\sim 0.45$ 
cts/s) and lower flux states (down to $\sim0.11$ cts/s).  We extracted a single spectrum 
from one of each of the three pointings.  An absorbed power-law\footnote{\correc{Note 
that we chose to use a simple power-law rather than the more sophisticated models usually 
used to fit CV spectra in order to compare the XRT spectral parameters to those mentioned in the 
literature. In particular, \citet{gotz06} showed that the extrapolation at hard X-rays of 
spectrum obtained with {\it ROSAT} was compatible with the \integral/IBIS one.  A discussion of the emission
processes at work in CVs is beyond the scope of this paper.}} fits the data well in all 
cases (\chisq=1.19 for 89 dof, 1.29 for 14 dof and 1.26 for 98 dof, for pointings \#1, 
2 and 3, respectively).  The best spectral parameters of all three pointings are reported in 
Table~\ref{tab:spectral}, and they are in good agreement with \correc{those obtained 
by \citet{gotz06} from a {\it ROSAT} observation of TW Pic.  In addition, no cut-off 
is seen in the XRT spectrum (which extends to higher energy than the {\it ROSAT} 
spectrum).  The extrapolation of the XRT spectral model to the 20--40 keV range 
leads to a flux that is compatible with the flux measured by \integral\ (0.9 mCrab). 
All these points (including the spatial coincidences discussed above) further confirm 
that IGR~J05346$-$5759 is TW Pic, including the the spectral 
variability of  IGR~J05346$-$5759 as TW Pic is known to be variable.} This 
variability has been used by \citet{norton00} to refute the Intermediate Polar (IP) 
type for this source. \correc{We therefore conclude that IGR~J05346$-$5759 is the 
hard X-ray counterpart to TW Pic, and thus, is a CV.}

\paragraph{\bf \object{IGR~J10101$-$5654: }\\}
A refined \correc{{\it Chandra}} position for this object has recently been published 
by \citet{tomsick08}.  The XRT position is \correc{0.55\arcsec\ from the 0.64\arcsec\ accurate
{\em Chandra} position \citep{tomsick08} and therefore both positions are compatible.}  
We further confirm all the suggested associations for 
this object, and the fact that it is a very likely HMXB \citep{masetti06c,tomsick08b}. There are 
no UVOT data available for this pointing.\\
\indent The spectrum is well-fitted with an absorbed power-law (C=19.9 for 14 bins). 
The spectral parameters reported in Table~\ref{tab:spectral} are fully consistent with 
those reported from the {\it Chandra} observation of this source \citep{tomsick08}. 
Although the poor statistical significance of the parameters we obtain does not allow 
us to constrain the possible spectral variability for this source, the flux we obtain 
from the \swift\ observation is about five times higher than during the {\it Chandra} 
observation \citep{tomsick08}. This may indicate significant variation of the mass
accretion rate. 

\begin{landscape}
\centering
\begin{table}

\caption{Magnitudes and UVOT position of the newly identified optical 
and UV counterparts in the USNO-B1.0 catalog (I, R and B bands) and \swift/UVOT detector (V, U, UVW1, UVM2, and UVW2  bands). The USNO-B1.0 photometric 
accuracy is typically 0.3 mag \citep{monet03}. The B magnitudes are those obtained from the USNO-B1.0
catalog, except where indicated. The long dashes indicate the absence of corresponding
data.}
\begin{tabular}{lccccccccccc}
\hline
\hline
Name & Optical counterpart & \multicolumn{2}{c}{UVOT position}  & \multicolumn{8}{c}{Magnitudes} \\
(IGR) &  (USNO-B1.0)       & RA      & DEC                  &  I & R & V & B & U & UVW1 & UVM2 & UVW2 \\
\hline
J03532$-$6829 & 0214-0026031 & 03h 52m 57.5s & $-$68\degr\ 31\arcmin\ 17.4\arcsec\ &12.7 & 12.3  & -- -- -- &13.7   & -- -- --  &-- -- -- &  -- -- -- & 17.28$\pm0.02$\\ 
J05346$-$5759 & 0319-0039890 & 05h 34m 50.6s & $-$58\degr\ 01\arcmin\ 40.8\arcsec\ &13.8 & 15.2  & -- -- -- &14.9   & -- -- --  & 13.886$\pm0.004$ & 13.182$\pm0.006$$^\ddagger$ & 12.909$\pm0.001$$^\ddagger$ \\
J09025$-$6814 & 0217-0159098$^\star$ & 09h 02m 39.5s & $-$68\degr\ 13\arcmin\ 38.2\arcsec\ &  -- -- -- &  8.6  &  -- -- -- &9.7 & 16.6 & 16.61$\pm0.02$$^\ddagger$ &17.63$\pm0.03$ & -- -- --  \\
J13000+2529   & 1154-0199710 & 12h 59m 55.3s & 25\degr\ 28\arcmin\ 10.5\arcsec\    & 10.6 & 11.3 &  -- -- -- &13.0   & -- -- --  & 15.51$\pm0.02$ & 17.61$\pm0.06$ & -- -- -- \\
J15161$-$3827  \#1 & 0515-0356635 &-- -- -- &-- -- -- & 10.7   & 10.6  &-- -- --  & 10.6   &-- -- -- &-- -- -- &-- -- -- & -- -- --\\
\hspace*{1.73cm}\#2 & 0515-0357047 &-- -- -- &-- -- -- & 18.2   & 18.3  & -- -- --  &19.0   &-- -- -- &-- -- -- &-- -- -- &-- -- -- \\
\hspace*{1.73cm}\#3 & 7822-02179-1 &-- -- -- &-- -- -- & 10.9   &  11.3   &-- -- --  &11.0  &-- -- -- &-- -- -- &-- -- -- & -- -- --\\
\hspace*{1.73cm}\#4 & 0515-0356459 &-- -- -- &-- -- -- &-- -- --  & 18.5  &-- -- --  &18.9 &-- -- -- &-- -- -- & -- -- --&-- -- -- \\
J15479$-$4529 &-- -- --  & 15h 48m 14.6s &-45\degr\  28\arcmin\ 39.9\arcsec &-- -- -- &-- -- -- &-- -- --  & -- -- --  &-- -- -- &-- -- -- & -- -- -- & 14.501$\pm0.003$$^\ddagger$\\
J17353$-$3539 &0543-0510755 &-- -- --               & -- -- --   &10.9  &   -- -- -- & 11.9  &-- -- --     & -- -- --  & $>$20.3 &$>$20.2  &-- -- -- \\
J17476$-$2253 &0671-0618341 &   -- -- --            & -- -- --   &15.3  & 17.0  &   -- -- --     &19.1   &-- -- --   & -- -- --  & $>$19.3   &-- -- -- \\
J18214$-$1318 &-- -- --  & -- -- --               &        -- -- --                          &-- -- -- &-- -- -- &$>19.3$& $>19.8$$^\diamond$ &$>19.9$ & $>20.6$ & $>20.5$ &$>20.9$ \\
J19267+1325   &-- -- --  & 19h 26m 27.0s & 13\degr\ 22\arcmin\ 05.1\arcsec &-- -- -- &-- -- -- &-- -- -- &-- -- -- & -- -- --& -- -- --& -- -- -- &20.54$\pm0.07$ \\
J20286+2544 \#1$^\dagger$ & 1157-0462303$^\star$ & 20h 28m 35.1s & 25\degr\ 43\arcmin\ 59.5\arcsec\ & -- -- -- & 10.1 & 15.06$\pm0.01$$^\ddagger$&11.4  & 18.03$\pm0.05$$^\ddagger$ & 20.5$\pm0.1$$^\ddagger$ & $>21.1$& 20.6$\pm0.1$$^\ast$  \\
\hspace*{1.73cm}\#2 & 1157-0462166         & 20h 28m 28.9s & 25\degr\ 43\arcmin\ 24.6\arcsec\ & 8.9 & 8.7  &  12.897$\pm0.007$$^\ddagger$ & 10.3 &15.41$\pm0.01$$^\ddagger$& 16.83$\pm0.02$$^\ddagger$& $>21.1$ & 19.15$\pm0.05$$^\ddagger$  \\
J23206+6431 & 1545-0296864 &-- -- -- &-- -- -- & 17.9 & 19.1 &  -- -- -- &20.9 &$>21.1$&-- -- -- &-- -- -- &-- -- -- \\
\hline
\label{tab:uvcounterparts}
\end{tabular}
\begin{list}{}{}
\item[$^\ddagger$]Values averaged over multiple pointings.
\item[$^\star$]There are two possible USNO-B1.0 sources in the XRT error circle. This is the closest to the IR source. 
\item[$^\dagger$]The UVOT positional accuracy is dominated by  a statistical uncertainty of 1.1\arcsec.
\item[$^\diamond$]B magnitude obtained from \swift/UVOT.
\item[$^\ast$]Average value obtained with {\tt{uvotsource}}.
\end{list}
\end{table}
\end{landscape}
\paragraph{\bf \object{IGR~J13000$+$2529: }\\}
Based on the spatial coincidence between the two objects, \cite{bassani06} suggested 
an association of IGR~J13000$+$2529 with \object{MAPS-NGP O-379-0073388}, an AGN listed
in the NED database.  The XRT position is consistent with that of MAPS-NGP O-379-0073388, 
which \correc{provides further confirmation that} the high energy source and the AGN are 
the same.  We found a single 2MASS source within the XRT error circle, and although the 
source is not reported as extended it lies only 0.9\arcsec\ from the position of the AGN 
reported in NED, which indicates the two objects are probably the same. \correc{A single 
source is also found within the XRT error circle in the USNO-B1.0 catalog and UVOT images 
(Table~\ref{tab:uvcounterparts}).}\\
\indent As the source is very weak, we extracted an average spectrum from the two
\swift\ pointings. The spectrum has too few counts for a spectral analysis to be possible.
Although this source is the faintest from our sample that we detect with XRT, and the 
very low flux could indicate a lower probability that it is associated with the IGR source, 
the good spatial coincidence with the AGN along with \correc{the fact that this is the 
only XRT source in the IBIS error circle that we detect make IGR~J13000$+$2529 a strong 
AGN candidate}.

\begin{table*}[htbp]
\caption{X-ray spectral analysis. \correc{Errors and upper limits are all given at 
the 90\% level.}}\label{tab:spectral}
\begin{tabular}{lcllll}
\hline\hline
Name & Net number & Galactic \nh\ (LAB/DL)$^\ddagger$ & \nh\ & $\Gamma$& 2--10 keV flux\\
(IGR) &  of counts & $\times10^{22}$~cm$^{-2}$ & $\times10^{22}$~cm$^{-2}$  &   & \ergcms \\
\hline
J03184$-$0014 & 19 & 0.05/0.06 & 0.06$^\dagger$ & 1.4$_{-0.7}^{+0.8}$ & 5.3$_{-0.3}^{+0.5}$ $\times10^{-14}$  \\
J03532$-$6829 & 1650 & 0.06/0.06 & 0.09$_{-0.04}^{+0.04}$ & 1.9$_{-0.1}^{+0.1}$ & 1.75$_{-0.18}^{+0.14}$ $\times10^{-11}$\\
J05319$-$6601 & 19 & 0.12/0.06 & 0.12$^\dagger$ & 1.55$_{-0.77}^{+0.89}$ & 5$_{-3}^{+4}$ $\times10^{-14}$\\
J05346$-$5759 & 2172 & 0.04/0.05 & $<$0.05 & 1.22$_{-0.09}^{+0.1}$ & 1.7$_{-0.1}^{+0.2}$ $\times10^{-11}$\\
              & 378  & & $<$0.15 & 1.75$_{-0.3}^{+0.3}$ & 5.7$_{-1.0}^{+1.2}$ $\times10^{-12}$\\
              & 2516 & & 0.05$_{-0.03}^{+0.03}$ & 1.34$_{-0.09}^{+0.09}$ & 1.7$_{-0.1}^{+0.1}$ $\times10^{-11}$\\
J09025$-$6814 & 17 & 0.05/0.07 & 9$_{-7}^{+123}$& $<$3.2 & $<9.2\times10^{-12}$\\
J10101$-$5654 & 86 & 1.35/1.77 & 3.3$_{-1.7}^{+2.5}$& 1.3$_{-0.8}^{+0.9}$ & 1.2$_{-0.6}^{+0.3}$ $\times10^{-11}$\\
J13020$-$6359 & 337 & 1.40/1.53  & 2.48$^\dagger$ & 0.9$_{-0.3}^{+0.3}$ & 2.3$_{-0.9}^{+0.3}$ $\times10^{-11}$\\
              & 670 & & 2.48$^\dagger$ & 1.2$_{-0.2}^{+0.2}$ & 2.6$_{-0.4}^{+0.2}$ $\times10^{-11}$\\
              & 471 & & 2.48$^\dagger$ & 1.1$_{-0.2}^{+0.2}$ & 2.3$_{-0.5}^{+0.3}$ $\times10^{-11}$\\
              & 574 & & 2.48$^\dagger$ & 1.1$_{-0.2}^{+0.2}$ & 2.3$_{-0.5}^{+0.3}$ $\times10^{-11}$\\
J15161$-$3827 \#1 & 48 & 0.06/0.07 & $22_{-9}^{+17}$ & 2.0$^\dagger$ & 1.2$_{-0.5}^{+0.5}$ $\times10^{-12}$\\
             \#2  & 32 & 0.07/0.07 & $<0.2$ & 1.2$_{-0.5}^{+0.7}$ & $<$1.3$\times10^{-13}$\\
             \#3  & 18 & 0.07/0.07 & $<1.9$ & $>2.8$ & $<$1$\times10^{-13}$\\
             \#4  & 13 & 0.06/0.07 &  0.065$^\dagger$ & 2.0$_{-0.9}^{+1.0}$ & 3$_{-2}^{+5}$ $\times10^{-14}$\\
J16287$-$5021 & 75 & 1.37/1.55 & 2.6$_{-1.6}^{+2.1}$& 0.9$_{-0.8}^{+0.8}$ & 6.5$_{-3.0}^{+2.2}$ $\times10^{-12}$\\
J17353$-$3539 & 416 & 0.69/0.63 & 0.7$_{-0.3}^{+0.4}$ & 2.2$_{-0.4}^{+0.4}$ & 5.0$_{-0.5}^{+0.9}$ $\times10^{-12}$\\
              & 803 & & 0.8$_{-0.2}^{+0.2}$ & 2.1$_{-0.3}^{+0.3}$ & 1.2$_{-0.1}^{+0.1}$ $\times10^{-11}$\\
J17476$-$2253 & 45  & 0.30/0.38 & 1.9$_{-1.1}^{+1.7}$ & 2.6$_{-1.0}^{+1.4}$ & 5$_{-3}^{+2}$ $\times10^{-12}$\\
J18214$-$1318 & 1866 & 1.21/1.54 & 3.5$_{-0.5}^{+0.8}$ & 0.4$_{-0.2}^{+0.2}$ & 6.7$_{-0.4}^{+0.7}$ $\times10^{-11}$\\
J19267+1325   & 461 & 0.95/0.93 & $<0.6$ & 1.1$_{-0.3}^{+0.3}$ & 8.1$_{-0.7}^{+1.6}$ $\times10^{-12}$\\
J20286+2544  \#1 & 171  & 0.20/0.26 & 61$_{-20}^{+23}$ & 2.5$_{-1.4}^{+1.6}$ & 2.1$_{-1.2}^{+1.6}$ $\times10^{-11}$\\
            \#2 & 53 & 0.20/0.26 & 93$_{-61}^{+80}$ & 2.7$_{-3.1}^{+3.1}$ & $<$1.6$\times10^{-11}$\\
J23206+6431   & 244 & 0.78/0.90 & 0.9$_{-0.7}^{+1.0}$ &1.6$_{-0.5}^{+0.7}$ & 5.5$_{-1.0}^{+1.3}$ $\times10^{-12}$\\
\hline
\hline
\end{tabular}
\begin{list}{}{}
\item[$^\ddagger$]Values of weighted average Galactic \nh\ respectively obtained from Leiden/Argentine/Bonn (LAB)  
and Dickey \& Lockman (DL) surveys of Galactic H~I in the Galaxy.
\item[$^\dagger$]Unconstrained parameter that was fixed during the spectral fit.
\end{list}
\end{table*}

\paragraph{\bf \object{IGR~J13020$-$6359: }\\}
This source was first mentioned in \citet{bird06} and was classified as a pulsar/HMXB in 
\citet{bird07}, probably based on the positional coincidence with \object{2RXP J130159.6$-$635806}, 
which indeed is an HMXB containing a pulsar \citep{chernyak05}. \citet{bodaghee07} further
report a distance to the source of about 5.5~kpc.  We find a single XRT source within the IBIS 
error circle at a position compatible with that of 2RXP J130159.6$-$635806. This renders the 
association even more likely.  It is unfortunate that due to its off-axis position (the 
pointings were aimed at PSR B1259$-$63), none of the UVOT exposures contains the source. 
There is no USNO-B1.0 source within the \swift\ error circle. We estimate a lower limit 
 V$\gtrsim$21 for the magnitude of an optical counterpart.
\correc{\citet{chernyak05} mention the presence of a J$\sim13$, H= 12.0 and 
K$_\mathrm{s}$=11.3 2MASS source at a position compatible with that of the pulsar, that they 
consider as its likely counterpart.}\\
\indent As the source may be significantly variable \citep{chernyak05}, we fitted each 
spectrum from each independent pointing separately. An absorbed power-law fits all spectra 
rather well (\chisq\ in the range 0.6 to 1.40 for 30 to 13 dof). Since the absorption is 
poorly constrained and given that \citet{chernyak05} mention a relatively stable value of 
2.48$\times10^{22}$~cm$^{-2}$, we froze \nh\ to this value in all our fits.  Note that for 
all pointings the value obtained for \nh\ when it is allowed to 
vary is in good agreement, or compatible with \citet{chernyak05}.  The spectral results reported in 
Table~\ref{tab:spectral} show some slight variability especially between the first 
pointing and the following ones, which are slightly softer.  The spectral parameters are 
those expected for an accreting pulsar and, assuming a distance of 5.5 kpc, lead to a 
2--10 keV luminosity of about 8--9$\times10^{34}$~erg/s, typical for these objects.

\paragraph{\bf \object{IGR~J15161$-$3827: } \\}
Based on the positional coincidence of IGR~J15161$-$3827 and \object{LEDA~2816946}, 
\citet{masetti06b} suggested that the latter, an AGN, is the counterpart of the high 
energy source. The AGN type is intermediate between a Liner and a Sey 2 at $z$=0.0365 
\citep{masetti06b}.  The \swift\ mosaic image revealed four possible X-ray counterparts 
within the IBIS error circle of IGR~J15161$-$3827. \object{Swift J151559.3$-$382548}, 
\object{Swift J151630.0$-$382656}, \object{Swift J151612.2$-$383102}, and
\object{Swift J151545.8$-$382738} are labeled source \#1, \#2, \#3, and \#4, respectively
in Tables~\ref{tab:position} and \ref{tab:ircounterparts}.  It is {\it a priori} not 
possible to say which (if any) is the true counterpart.  Two of these are compatible with 
IR counterparts found in the 2MASS and 2MASX catalogs, although 2MASX J15155970$-$3825468 
\correc{is 4.9\arcsec\ from the \swift\ position and therefore is} slightly outside the 
XRT error circle of source \#1. \correc{It is, however, an extended source, and the XRT
error circle still contains a significant part of the source.}  This source is the one 
suggested by \citet{masetti06b} as the counterpart to the IGR source. \correc{A USNO-B1.0 
source lies at 5.4\arcsec\ from the XRT position, at a position compatible with the 2MASX 
source (offset by 0.7\arcsec), given the extension of the latter.}  Source \#3 has a 
position compatible with an IR point source, which is consistent with being TYC 7822-2179-1 
catalogued as a star in SIMBAD \correc{and also reported in the USNO-B1.0 catalog 
(Table~\ref{tab:uvcounterparts}). There are USNO-B1.0 counterparts for the other two 
sources as well, although the source \#4 counterpart does not have measurement in the 
I-band (Table~\ref{tab:uvcounterparts}).}  There are no UVOT data available for either 
of the two XRT pointings. \\
\begin{figure}[htbp]
\epsfig{file=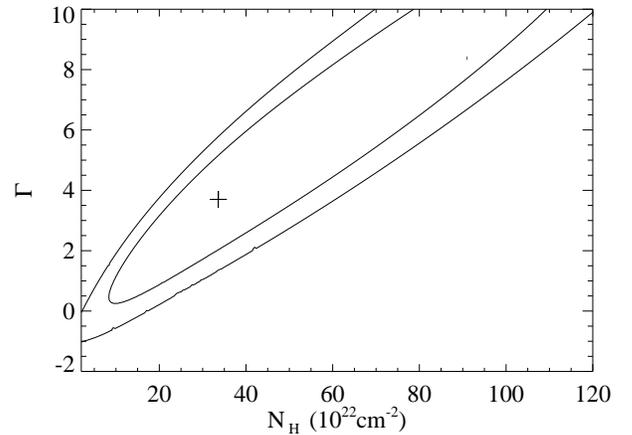,width=\columnwidth}
\caption{Contour plot of the power-law photon index
$\Gamma$ vs \nh\ in IGR~J15161$-$3827 source \#1. The contours 
represent $\Delta C$=2.30 and 4.61.}
\label{fig:nh}
\end{figure}
\indent We extracted an average spectrum from the two pointings for each of the four sources.
\correc{The spectrum of source \#1 has a low statistical quality. The spectrum was 
fitted with an absorbed power-law (C=38.5 for 15 bins).}  When all parameters are left 
free to vary, they are very poorly constrained \correc{(Table~\ref{tab:spectral}). Although 
only an upper limit can be obtained from the absorption, visual inspection of the spectrum 
shows that the source may show significant absorption. 
Fig.~\ref{fig:nh} represents the contour plot of $\Gamma$ vs. \nh. It is clear from this 
figure that the value of \nh\ is tightly correlated to that of $\Gamma$ as expected.
This figure, however, shows that for $\Gamma\geq$0.5, a value typical for most high energy 
sources, this source is significantly (intrinsically) absorbed} as would be expected from a Sey 2. 
\correc{We remark that, to obtain the 20--40 keV flux of 0.5 mCrab seen with \integral\
\citep{bird07}, a harder power-law ($\Gamma\sim0.7$) is needed.}  Even in that case, 
significant absorption is implied by the fit.  The 2--10 keV luminosity at $z$=0.0365 is 
5.6$\pm0.5 \times10^{42}$ erg/s, compatible with the luminosity of an AGN.\\
\indent \correc{An absorbed power-law provides a good fit to the spectrum of source \#2 
(C=7.6 for 15 bins). The spectrum is consistent with little or no absorption in 
this source.}  The absence of significant absorption in the spectrum of the source argues 
in favour of a nearby object.  The extrapolated 20--40~keV flux is well below the 
\integral\ flux. \correc{A hard power-law with a photon index $\lesssim0.35$ would be 
needed to reach the 20--40 keV flux observed by \integral.} These last points argue 
against an association of source \#2 with the IGR source.\\
\indent The X-ray spectrum of source \#3 is well-fitted with an absorbed power-law 
(C=7.7 for 14 bins).  The quite steep power-law and the low flux obtained with 
the lower limit of $\Gamma$, may indicate that the spectrum is thermal. Replacing 
the power-law by a black-body also gives a good fit (C=7.5 for 15 bins).  
\correc{Note that since the value of \nh\ is poorly constrained, it was frozen to the 
value of Galactic \nh.}  The black-body temperature is 0.2$\pm0.1$~keV for a luminosity 
of 9$\times(D_{10}^2)\times10^{32}$~erg/s, with $D_{10}$ the distance in units of 
10 kpc. The \correc{probable} low value of the absorption 
and the bright IR and optical counterparts argue in favour of a nearby object. In that 
case, the rather flat SED, black-body shape and temperature of the X-ray spectrum  
indicate that this is probably a young stellar object (YSO), e.g. a T Tauri star. 
 \correc{The softness of the source renders it difficult to reconcile the emission of 
this object with that at energies $>$20 keV. A very hard photon index of $\sim 1.0$ 
would be needed to be compatible with the 20--40 keV flux. Such a power-law slope is 
incompatible with the XRT spectrum.} We conclude that this object is certainly not 
related to the IGR source.\\
\indent  As for the 2 previous objects, the X-ray spectrum of source \#4 is well-fitted 
with an absorbed power-law (C=4.7 for 15 bins). \correc{A quite absorbed source 
with a very steep power-law seems to be favoured here.  We note, however, that a simple 
power-law (with no absorption) leads to more physical results for this source. As a 
compromise the value of absorption was frozen to the Galactic \nh.} A 0.6$_{-0.2}^{+0.3}$ 
keV black-body also fits the data well (C=6.2 for 15 bins). In any case, the 
extrapolation of the spectra to the \integral\ range \correc{falls well below the 
20--40 keV flux. A power-law with a value of the photon index incompatible with the 
XRT spectrum ($\Gamma\lesssim0.5$) would be needed.} This shows that this source 
and the IGR source are very probably not related.\\
\indent To conclude, the broad band (counterpart and X-ray) analysis of the four 
\swift\ objects found within the IBIS error circle of IGR~J15161$-$3827 leads us 
to conclude that the IGR source is very probably associated with the Liner/Sey 2
object LEDA~2816946.
\paragraph{\bf \object{IGR~J15479$-$4529: } \\}
Based on the presence of a {\it ROSAT} source (also detected by {\it{XMM-Newton}}) 
within the IBIS error circle, \citet{tomsick04} suggested an association between 
\object{1RXS J154814.5$-$452845}, \correc{and the IGR source.} 1RXS J154814.5$-$452845 
is a CV, more precisely an Intermediate Polar (IP) with a pulse period of 693~s and 
an orbital period of 562~min \citep{barlow06}. The refined position we obtained with 
\swift\ is only 5\arcsec\ \correc{from the {\it ROSAT} position \citep{haberl02}, 
indicating that the two positions are compatible}. \correc{There is a single source 
listed in SIMBAD within 3\arcmin\ of the XRT position.  This source has several 
names, one of which is \object{V~$^\star$ Ny Lup} indicating that it is a variable 
star \citep{samus04}}. \correc{Clearly the coincidence of the \swift\ and {\it ROSAT} 
sources renders their association likely. The fact that it is an IP, which are known 
hard X-ray emitters, strengthens the associations with the \integral\ source.}
We therefore confirm all suggested association, and the fact that IGR~J15479$-$4529 
is very probably an IP.  A bright source is found within the XRT error circle with 
the UVOT UVW2-filter (Table~\ref{tab:uvcounterparts}). \correc{Its position is 
consistent with the 2MASS source.}  We note that this UV counterpart may show some 
variability from one pointing to the other, from UVW2=14.0 to 15.0, \correc{which 
further confirms the variable nature of the source.}\\
\indent As the source may show some variability, we extracted a spectrum from each 
of the six pointings. Pointings \#1 and \#3 are quite short ($<1$~ks) so we do not 
consider them further. An unabsorbed power-law provides \correc{acceptable} fits to 
pointings \#2 and 4 (\chisq\ between 1.3 for 61 dof and 1.6 for 95 dof), but not to 
pointings \#5 and \#6, where a significant excess is detected at soft X-rays. 
\citet{haberl02} also mention the need for a black-body to account for a soft excess 
in their {\it XMM} spectra.  Adding a black-body to the power-law improves the fits 
greatly. We point out that \citet{haberl02} used a much more sophisticated model, 
but given the lower quality of our data, we only use the simple phenomenological 
models.  However, since they report some absorption in the spectra we also included 
an absorption component.  \correc{The resulting model is therefore 
{\tt{phabs*(bbody+powerlaw)}} in the {\tt{XSPEC}} terminology.}  When left free to 
vary, \nh\ tends toward very low values, although the 90\% upper limit is (marginally) 
compatible with $\sim0.14\times10^{22}$~cm$^{-2}$ \citep{haberl02}.  We therefore fixed
\nh\ to this value in our fits.  The results are reported in Table~\ref{tab:spec15479}. 
The variations of the flux do not seem to be related to spectral changes, but they are 
more probably due to slight variations of the accretion rate. 

\begin{table}[htbp]
\caption{Spectral parameters obtained from the fits to the XRT spectra of IGR~J15479$-$4529.
The model consists of \correc{black-body emission and a power-law, both modified by absorption}.}
\begin{tabular}{ccccc}
\hline
\hline
Pointing & kT$_{bb}$ & $\Gamma$ & \chisq & Flux\\
\#       & (keV) &  & (dof)& (\ergcms)\\
\hline
2        &0.12$_{-0.02}^{+0.03}$ &  0.9$_{-0.15}^{+0.06}$ &  1.0 (59) & 2.1$_{-0.2}^{+0.2}$ $\times10^{-11}$\\ 
4        &0.12$_{-0.01}^{+0.01}$ &  0.89$_{-0.09}^{+0.1}$ &  1.0 (93) & 2.8$_{-0.2}^{+0.2}$ $\times10^{-11}$\\ 
5        &0.11$_{-0.01}^{+0.01}$ &  0.9$_{-0.1}^{+0.1}$ &  1.1 (58) & 3.3$_{-0.3}^{+0.3}$ $\times10^{-11}$\\ 
6        &0.13$_{-0.01}^{+0.02}$ &  0.8$_{-0.2}^{+0.2}$ &  0.8 (43) & 3.4$_{-0.5}^{+0.4}$ $\times10^{-11}$\\ 
\hline
\end{tabular}
\label{tab:spec15479}
\end{table}

\paragraph{\bf \object{IGR~J18214$-$1318: } \\}
\correc{\citet{tomsick08} recently reported a refined X-ray position with 
{\it Chandra} for this object. The accuracy of their position is 
0.64\arcsec. The XRT position is 1.1\arcsec\ away from the {\it Chandra}
position, and the XRT error box (Table~\ref{tab:position}) contains the Chandra source.}
 \correc{No counterpart 
is detected in any of the UVOT filters.} We refer to \citet{tomsick08} for 
the identification of counterparts. An absorbed power-law fits the XRT 
spectrum well (\chisq=0.96 for 83 dof).  The value of \nh\ is higher than 
the Galactic value along the line of sight \correc{(Table~\ref{tab:spectral}), 
which confirms that there is intrinsic absorption in this source 
\citep{tomsick08}.}  Our value of 3.5$\times10^{22}$ cm$^{-2}$ is, however, 
significantly lower than the value of \correc{11.7$\times10^{22}$~cm$^{-2}$} obtained with {\it Chandra} 
observations \citep{tomsick08}. Fixing \nh\ to the latter value does not 
lead to a good fit (\chisq=2.4 for 84 dof). This indicates that the variations 
of \nh\ are genuine for this source. This further argues in favour of an HMXB 
(possibly a supergiant system) since significant variability of \nh\ has been 
reported for several systems \citep[e.g.,][in the case of IGR~J19140+0951]{prat08}.
Note that the very hard spectrum may then indicate the presence of a pulsar.

\paragraph{\bf \object{IGR~J19267$+$1325: }\\}
No X-ray source is found within the 3.7\arcmin\ IBIS error circle.  A bright 
X-ray source is, however, found 4.5\arcmin\ away from the center of the IBIS
error circle, and is, therefore, marginally compatible (within the 3$\sigma$ 
error circle) with the \integral\ position.  The \swift\ position is 
\correc{1.7\arcsec away} from the very recent \correc{0.64\arcsec} 
{\it Chandra} position reported by \citet{tomsick08b}. \correc{The positions 
given by the two satellites are therefore entirely compatible.}  \citet{tomsick08} 
report the presence of a single IR and optical counterpart within the {\it Chandra} 
error circle of this object.  We detect a single source in the UVOT detector 
\correc{(Table~\ref{tab:uvcounterparts}). It is well within the XRT and 
{\it Chandra} error circles (at 0.3\arcsec\ from the best {\it Chandra} position).}\\
\indent  \correc{An absorbed power-law provides an acceptable, although not perfect, 
fit} (\chisq=1.7 for 18 dof) to the XRT spectrum. The value of the absorption is 
below the Galactic value on the line of sight, and we obtain an upper limit 
consistent with the value \correc{of 2.1 $\times10^{22}$~cm$^{-2}$} obtained with 
{\it Chandra} \citep{tomsick08b}. \citet{landi07} mentioned the presence of 
black-body emission in the spectrum. We added such a component in our spectral 
fits (both with and without absorption), but in no case did it provide a 
noticeable improvement over the absorbed power-law fit. The extrapolated 20--40 keV 
flux of $\sim$2.3$_{-1.1}^{+1.7}$ mCrab is higher than the IBIS 20--40 keV 
flux of 0.7 mCrab reported by  \citet{bird07}.  This may argue in favour of an association of this source with 
the \integral\ source, suggesting that it undergoes significant flux variations. 
The hard power-law index, low value of the absorption and position on the plane 
of the sky close to the Sagittarius arm would tend to suggest this object has a 
Galactic origin.  \correc{Optical 
observations allowed \citet{steeghs08} to detect a possible counterpart within the 
{\em{Chandra}} error box of this source. Optical spectroscopy of this source 
permitted \citet{steeghs08} to further conclude that this source is a CV, probably
containing a magnetic white dwarf (see also Butler et al., submitted to ApJ).}

\begin{figure*}
\centering
\epsfig{file=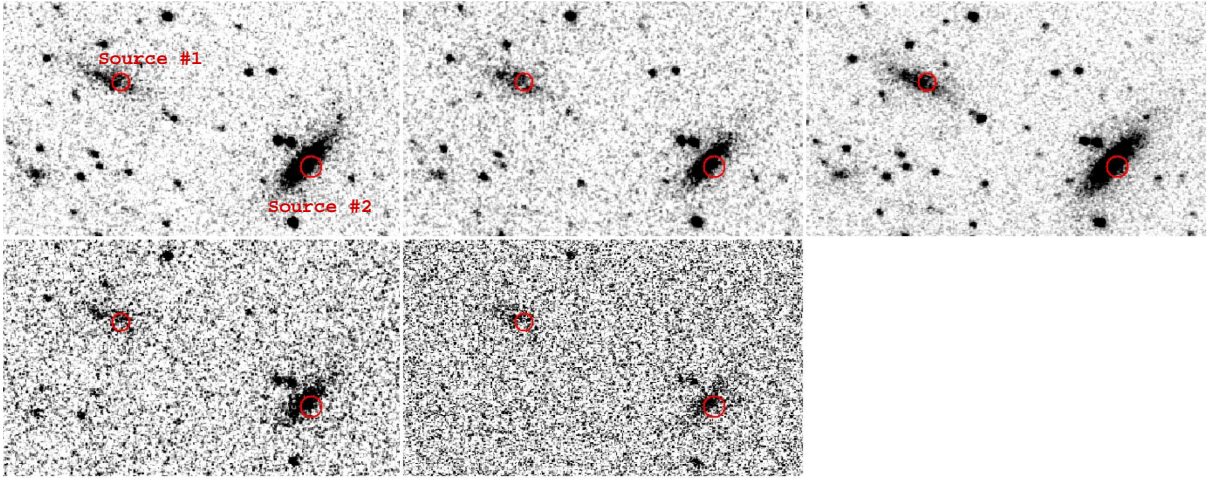, width=16cm}
\caption{From top to bottom and left to right 2.9\arcmin$\times$1.7\arcmin\  B, U, V, UVW1, UVW2  
images of the field of IGR~J20286$+$2544. The circles represent the \swift\ error circles 
for the two possible counterparts.}
\label{fig:20286}
\end{figure*}

\paragraph{\bf \object{IGR~J20286$+$2544: }\\}
Based on the presence of \object{MCG+04-48-002} in the IBIS error circle of the 
\integral\ source \citet{bassani06} suggested an association between the two
objects. \citet{masetti06a} added that although this Compton thick $z$=0.013 
Sey 2 was most probably the true counterpart to the IGR source, contribution 
from the nearby $z$=0.01447 galaxy NGC 6921 could not be excluded. Our \swift\ 
mosaic image reveals \correc{2 sources (\object{Swift~J202834.9+254359}, source 
\#1, and \object{Swift~J202828.7+254322}, source \#2), whose positions match 
those of the two galaxies.} \correc{There are two possible USNO-B1.0 sources 
within the \swift\ position of source \#1. Only one has well-estimated magnitudes 
in the B and R bands.  As it is the closest in position to the 2MASX source 
(0.9\arcsec), it is the one we report in Table~\ref{tab:uvcounterparts}.}
Both sources are quite well-detected with the UVOT as extended sources in the 
B, U, V, UVW1 and UVW2 filters (Fig.\ref{fig:20286}). \correc{The UVOT counterpart 
to source \#1 is not spontaneously found by {\tt{uvotdetect}}, although it is 
clearly visible in Fig.~\ref{fig:20286}.  In this case, we used {\tt{uvotcentroid}} 
to obtain an {\it{estimate}} of the source position\footnote{{\tt{uvotcentroid}} 
obtains mean coordinates by running a series of Monte-Carlo simulations
of the source's pixel distribution on a 20$\times$20\arcsec\ sub-image centred 
on the best position}, while the magnitudes at the best position of the source 
were obtained with {\tt{uvotsource}}}.  \correc{The positions of all counterparts 
of source \#2 are compatible with being within the extension of the 2MASX sources. 
We note, however, a large discrepancy between the B magnitude obtained by the 
UVOT (14.3) and that of the USNO-B1.0 source reported in Table~\ref{tab:uvcounterparts}. 
This may indicate that all UVOT magnitudes are over-estimated, possibly because 
of the extension of the source.}\\
\begin{figure*}
\centering
\epsfig{file=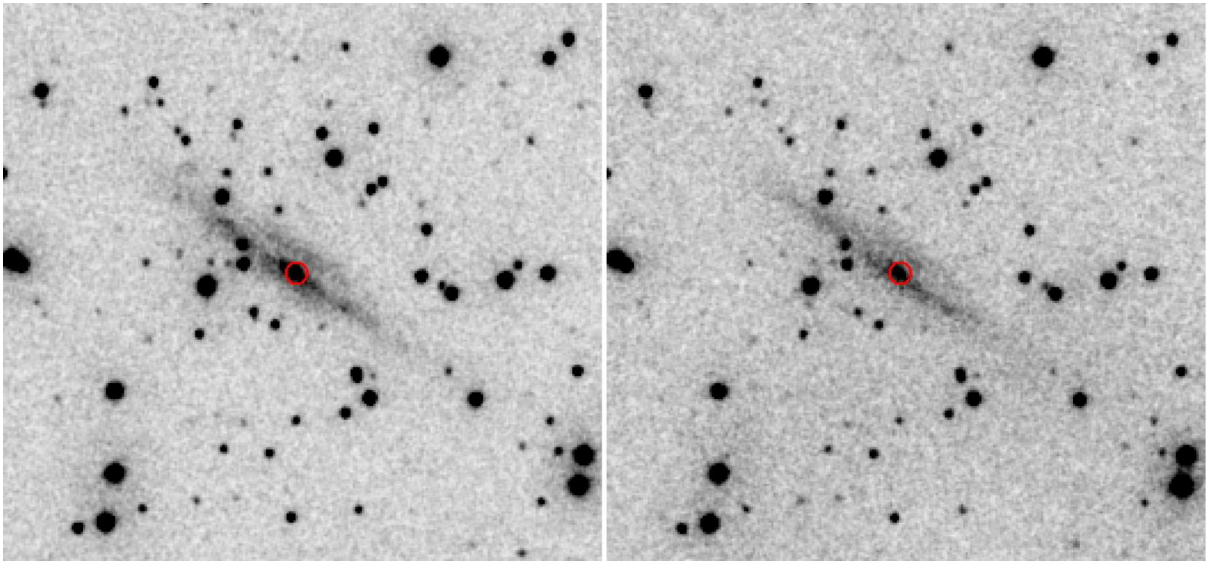,width=16cm}
\caption{4.3\arcmin$\times$4.1\arcmin\ U (left) and UVW1 (right) UVOT images of the field around 
IGR~J09025$-$6814. The best X-ray position is represented by the circle.}
\label{fig:09025}
\end{figure*}
\indent As both sources are rather faint, we accumulated average spectra from the four
pointings. The spectrum of source \#1 is not well-fitted by an absorbed power-law 
(C=43 for 14 bins). Significant residuals are found at low energy. Such soft 
excesses have been reported in a number of AGN (e.g., paper 1 and references therein). 
Adding an unabsorbed black-body greatly improves the fit (C=8.0 for 14 dof). 
The black-body has a temperature of 0.4$_{-0.1}^{+0.2}$ keV, and a 0.5--10 keV 
luminosity of 1.5$_{-0.5}^{+0.75}$~$\times10^{40}$ erg/s assuming a distance $z$=0.013. 
The other parameters are reported in Table~\ref{tab:spectral}.  The source is strongly 
absorbed, but not Compton-thick.  The extrapolated 20--40 keV flux is 4.5 times lower 
than the 20--40 keV IBIS flux \correc{of 2.6 mCrab} reported by \citet{bird07}.\\
\indent  As for source \#1, a simple absorbed power-law does not provide a good description of 
the spectrum of source \#2. It in particular gives negative values for the power-law index. 
Even fixing the latter to a fiducial value of 2 does not help. We used a similar model as for 
source \#1, and 
this led to a good fit (C=11.7 for 14 bins). The value of the photon index is 
poorly constrained (Table~\ref{tab:spectral}). In subsequent runs it was fixed to 
2.0. \correc{Even in those cases, the source is highly absorbed and could be a 
Compton-thick object with \nh$\sim83\times10^{22}$~cm$^{-2}$.  In this latter case, 
the extrapolated 20--40 keV flux is 8.2 times} lower than the IBIS flux of 
IGR~J20286$+$2544. \\
\indent Although the flux of source \#2 highly depends on the value of the photon 
index, our results indicate that IGR~J20286$+$2544, the source seen by \integral, 
is probably a blend between Swift~J202834.9+254359 and Swift~J202828.7+254322, with 
a stronger contribution from the former. \correc{We also note that the high flux 
obtained by \integral\ may indicate significant variability in those sources.} 
It has to be noted that the high absorption in source \#2 would argue in favour
of the source being a Sey 2, similar to source \#1.

\paragraph{\bf \object{IGR~J23206+6431: }\\}
This source was associated with \object{2MASX~J23203662+6430452} by \citet{bikmaev08} 
based on the observation made with \swift. They did not provide any fine X-ray 
position, however. The position reported in Table~\ref{tab:position} is fully 
compatible with that of the IR counterpart. They measured a value of $z$=0.0717 
from optical spectroscopy of this counterpart, and classified it as a Sey 1. The 
source is not detected by the UVOT U-filter with \correc{a 3$\sigma$ lower limit} 
U$>$21.1.\\
\indent An absorbed power-law fits the spectrum well (\chisq=0.3 for 8 dof). 
The 2--10 keV luminosity at $z$=0.0717 is 5.4$_{-1.0}^{+1.3}\times10^{43}$~erg/s, 
which is typical for this type of object.  The low value of the absorption is 
also compatible with the source being a Sey 1.

\subsection{\object{IGR~J03184$-$0014}}
The position of the \swift\ source we found is 4.4\arcmin\ away from the best IBIS 
position, and is, therefore, slightly outside the 4.0\arcmin\ 90$\%$ IBIS error 
circle reported in \citet{bird07}. Given the compatibility of the 3$\sigma$ error 
circles of both the \integral\ and \swift\ sources, we first consider the possibility
that the two sources are associated.  Its IR counterpart has a well-measured 
magnitude in the K$_s$ band only. There is no USNO-B1.0 source within the \swift\ 
error circle \correc{with V$\gtrsim21$}. The UVOT telescope observed the field in 
the UVW1 filter.  The {\tt{uvotdetect}} tool did not yield a detection of a source 
within the XRT error circle. The presence of a bright UVW1=13 source at 23.8\arcsec\  
from the candidate counterpart renders, however, the detection of a possible 
counterpart difficult (the source is so bright that part of its flux is within the 
XRT error circle).  Keeping this caveat in mind, we can roughly estimate a 3$\sigma$ 
upper limit UVW1$>21.95$ based on the faintest source detected (at a confidence level 
greater than 3$\sigma$) with {\tt{uvotdetect}}.\\
\indent The \swift\ spectrum extracted from the single pointing available has 24~cts. 
An absorbed power-law is a good representation of the spectrum (C=10.4 for 14 bins). 
As the value of the absorption is very poorly constrained ($<1.3\times10^{22}$~cm$^{-2}$ 
at 90\% confidence if left free to vary) we fixed it to the Galactic value along the 
line of sight. The spectral parameters are reported in Table~\ref{tab:spectral}. A 
fit with a black-body instead of the power-law also provides a good description of 
the data although statistically worse (C=12.0 for 14 bins). The black-body has 
a temperature of 1.0$_{-0.3}^{+0.7}$ keV, and a luminosity of 
1.5$_{-0.7}^{+1.5}\times 10^{33}$erg/s at a distance of 10~kpc. The extrapolated 
20--40 keV flux (3.5$\times 10^{-13}$\ergcms) is $\sim100$ times below the IBIS flux 
reported in \citet{bird07}. We, therefore, conclude that this source 
(\object{Swift J031818.0$-$001748}) and IGR~J03184$-$0014 are probably not related.\\
\indent Given the faintness of the source, it is quite difficult to unveil its true 
nature. The fact that it is well-detected in the K band only, and that it has no 
counterpart in the optical and UV bands either points to a very distant object or a 
faint Galactic source. If we assume the source is an AGN, with a luminosity of 
6$\times10^{42}$~erg/s (the luminosity of the faintest AGN detected in paper 1), this 
implies a distance $z$=0.144.  The only source that was farther than this in paper 1 
(IGR~J09523$-$6231 was not significantly detected in the IR, but had, on the other 
hand, a well detected U-band counterpart compatible with the emission from the 
accretion disc of the AGN. The absorption on the line of sight for the latter 
object was also much higher than in the case of IGR~J03184$-$0014, which suggests 
that, if IGR~J03184$-$0014 was an AGN it would probably be detected with the UVOT. 
We conclude that it is unlikely that this object is an AGN.  In the case of a 
Galactic object, the spectral parameters, while being very poorly constrained, may 
be compatible with the source being either an active star, a CV, or a neutron star 
X-ray Binary.  At 8~kpc, the 2--10 keV power-law luminosity would be 
1.1$\times10^{34}$erg/s.  These again point towards the \swift\ and \integral\ 
sources not being related.

\subsection{\object{IGR~J05319$-$6601}}
A weak source is found in the XRT $\sim$20~ks mosaic image at a position consistent 
with that of IBIS \citep{gotz06}. The XRT position is also consistent with that of 
\object{RX~J0531.8$-$6559}. There are no IR or optical counterparts reported 
in the 2MASS, 2MASX, USNO-B1.0 catalogs \correc{with K$_{s}\gtrsim16.2$, and 
V$\gtrsim21$}.  There are no sources detected in the UVOT U, V, UVM2 and UVW2 
filters compatible with the XRT position. As in the case of IGR~J03184$-$0014, 
the presence of a bright UV source  at $\sim10$\arcsec\ from the centre of the XRT 
error box renders the estimate of upper limits difficult due to possible contamination
at the position of IGR~J05319$-$6601. 
In a similar manner as for the previous source, we can estimate U$>19.43$, V$>19.33$, 
UVW1$>19.81$, and UVM2$>14.87$.\\
\indent An absorbed power-law is a good representation of the \swift\ spectrum 
(C=7.4 for 14 bins). As the value of the absorption is very poorly constrained 
($<2.7\times10^{22}$~cm$^{-2}$ at 90\% confidence if it is left free to vary), we 
fixed it to the Galactic value along the line of sight. The spectral parameters are 
reported in Table~\ref{tab:spectral}. A fit with a black-body instead of the power-law 
also provides a good description of the data (C=6.53 for 14 bins). The black-body 
has a temperature of 0.8$_{-0.3}^{+0.4}$ keV, and a luminosity 
6.6/$D_{10}^2$$_{-2.9}^{+4.5}\times 10^{32}$erg/s, \correc{where $D_{10}$ is the distance 
in units of 10 kpc}. The extrapolated 20--40 keV flux (based on the power-law model) 
is \correc{within 9.6$\times10^{-15}$--1.9$\times10^{-13}$~\ergcms, which is more than 
40 times} lower than the IBIS \correc{20--40 keV flux of 0.9 mCrab} reported in 
\citet{gotz06}. We remark, however, that during a second observing campaign, the same 
team did not detect the source with \integral, which may indicate significant variability. \\
\indent \citet{gotz06} suggested the IGR source may be an X-ray binary in the LMC. In fact 
this assumption is in good agreement with the fact that no counterparts are reported
in any of the optical and IR catalogs which may be due to the large distance to the source. 
Assuming the source is at the distance of the LMC, the 2--10 keV luminosity is  
1.6$_{-0.3}^{+0.5}\times10^{34}$~erg/s, which is therefore compatible with this hypothesis.

\subsection{\object{IGR~J09025$-$6814}}
A very weak XRT excess is found within the IBIS error circle. The XRT position contains 
a 2MASX source (Table~\ref{tab:ircounterparts}). \correc{It also contains two USNO-B1.0 
sources. The one that is reported in Table~\ref{tab:uvcounterparts} is the closest to 
the position of the 2MASX source (1.1\arcsec\ away).  It also has well-defined B and I 
magnitudes while the second source does not.} The 2MASX source is reported in the NED 
database as ESO 60-24/NGC 2788A, a $z$=0.013 galaxy. The detection of the source at 
X-ray energies with \integral\ and \swift\ suggests it is an AGN. The X-ray position 
falls right on the nucleus of the Galaxy as can be seen in the UVOT U and UVW1 images 
of the field (Fig.~\ref{fig:09025}).\\
\indent The XRT source is too weak to study any possible variability. We therefore 
extracted an averaged spectrum from the four pointings.  An absorbed power-law seems 
to be a good representation of the spectrum. If we allow all parameters to be free to 
vary, they are, however, very poorly constrained (C=23 for 14 bins, 
\nh$<52\times10^{22}$ cm$^{-2}$ and $-2.5 <\Gamma<3$). \correc{In order to try and 
have a more constraining range of values, we refitted the spectrum forcing $\Gamma\geq 0.$. 
An equally good fit is obtained with C=24 for 14 bins. The values are reported in 
Table~\ref{tab:spectral}.}  The source may be intrinsically absorbed, and  
this may point towards a Sey 2 object, 
as intrinsic absorption is expected in this case.  \correc{As the source 
is a Sey candidate, and to obtain a reasonable estimate of its flux, we fixed the power law 
photon index to 2.0. The 2--10 keV unabsorbed flux is 2.7$_{-1.5}^{+1.7}\times 10^{-12}$~erg~cm$^{-2}$~s$^{-1}$, 
which translate into a 2--10 keV luminosity of 8.7$_{-1.5}^{+1.7}\times10^{41}$ erg/s}. This value lies 
in the usual range for Seyfert galaxies.

\subsection{\object{IGR~J16287$-$5021}}
The XRT position is well within the \correc{4.4\arcmin} IBIS error circle, and is 
compatible with the very recent {\it Chandra} position reported by \citet{tomsick08b} 
\correc{(the {\it Chandra} positional accuracy is 0.64\arcsec. The \swift\ position is 
3.6\arcsec\ away from the {\it Chandra} position)}.  There are no infrared and optical 
counterparts reported in the 2MASS, 2MASX, USNO-B1.0 catalogs. There is no source 
within the XRT error circle in the UVOT UVM2-filter image with UVM2$>20.0$.\\
\indent The XRT spectrum is well-fitted by an absorbed power-law (C=8.5 for 
14 bins). The value of the absorption is \correc{not very well-constrained 
(Table~\ref{tab:spectral}), but may indicate little intrinsic absorption}. 
\correc{Following \citet{tomsick08b}, we also fitted the data with a non-absorbed 
power-law. The fit has a worse C-statistic value of 19.5 for 14 bins, which indicates 
that absorption is required in the fit.} A good fit is also obtained when fixing 
\nh\ to the Galactic value along the line of sight (C=9.15 for 14 bins).  The 
spectrum is then much harder ($0.4\pm0.4$) and is not consistent with the very 
hard photon index \correc{of $-0.9\pm0.4$} obtained with {\it Chandra}  
\citep{tomsick08b}. Such a hard spectrum may indicate that the source is an HMXB. 

\subsection{\object{IGR~J17353$-$3539}}
As for the previous sources, a single X-ray source is found within the 
\correc{$\sim3$\arcmin\ } IBIS error circle. Our best position is within 3.1\arcsec\ 
of \object{1RXH~J173523.7$-$354013}, indicating that the two sources are the same. 
Note that the position of 1RXH~J173523.7$-$354013 reported in SIMBAD is at 
$\sim9$\arcsec\ from the position reported in the online {\it ROSAT}
catalog\footnote{http://www.xray.mpe.mpg.de/cgi-bin/rosat/src-browser}.
In addition to the 2MASS source listed in Table~\ref{tab:ircounterparts}, the XRT 
error circle also contains \correc{two USNO-B1.0 objects. Both have positions that 
are compatible with the position of the IR source. The closest (at 0.2\arcsec\ from 
the 2MASS source) is the one reported in Table~\ref{tab:uvcounterparts}}. No source 
is detected in the UVM2 and UVW1 filters of the UVOT telescope.\\
\indent Since we see some variability, we extracted spectra from \correc{all pointings 
and analysed them separately. We report here only the two extreme cases, as the others 
have parameters that are intermediate between those two.}  An absorbed power-law fits 
both spectra well (\chisq=0.75 and 0.88 for 16 and 34 dof, respectively).  The value 
of \nh\ is consistent with the Galactic value on the line of sight, which indicates 
the object is not highly intrinsically absorbed. The position of the source towards 
the Galactic Bulge may indicate a Galactic source.  We note that the absence of a 
UV counterpart with the presence of a possible optical counterpart is also more 
compatible with a Galactic source as, in case of an AGN, the optical would be also 
completely absorbed, while a Galactic stellar component could have significant 
emission in optical and not in the UV domain (see, e.g., paper 1). The compatibility 
of \nh\ with the Galactic value may indicate that the source lies at a significant 
distance. The 2--10 keV luminosity of the highest state (Table~\ref{tab:spectral}) 
is \correc{14.4$\pm0.1$ /D$_{10}^{2}$ $\times10^{34}$~erg/s (where D$_{10}$ is the 
distance in units of 10 kpc)}, which, combined with the spectral shape, may 
indicate the source is an HMXB.

\subsection{\object{IGR~J17476$-$2253}}
A single bright X-ray source is found within the IBIS error circle. \correc{A single 
source is reported in the 2MASS catalog (Table~\ref{tab:ircounterparts}), while 2 
USNO-B1.0 sources are found in the XRT error circle. The latter two are at, respectively, 
1.7 and 2.9\arcsec\ from the 2MASS source, and we consider the first (reported in 
Table~\ref{tab:uvcounterparts}) as just marginally compatible. The second is very probably 
not related to the IR source.}  No source is found in the UVM2-filter image of the UVOT 
telescope.\\
\indent The XRT spectrum is well-fitted with an absorbed power-law (C=4 for 15 bins). 
The value of the absorption is not well-constrained, and it may indicate that some 
intrinsic absorption occurs in this source. \correc{We, however, note that it is 
marginally compatible with the Galactic value along the line of sight. Fixing \nh\ to 
the Galactic value also provides a good description of the spectrum (C=11.2 for 
15 bins). In this case, the photon index is harder ($\Gamma=1.2\pm0.4$). In this latter 
case,} the 20--40 keV extrapolated flux is in good agreement with the 20--40~keV 
\integral\ flux of 1.3 mCrab \citep{bird07}.  This may further argue in favour of an 
association between the \swift\ and \integral\ sources, \correc{although the flux 
obtained when all parameters are left free to vary is lower than that obtained with 
\integral.}  We, in addition, note that an absorbed black-body also gives a good 
representation of the data. It has a temperature of 0.9$_{-0.2}^{+0.1}$~keV and a 
luminosity of 6$\times10^{34}$~erg/s at 10~kpc.  \citet{bird07} tentatively classify this 
source as an AGN. We do not find strong evidence in favour of this possibility, 
as the spectral parameters are also compatible with a Galactic X-ray binary. Here again,
 the position towards the Galactic bulge may favour a Galactic source.  We note that 
the absence of a UV counterpart with the presence of a 
possible optical one is also more compatible with a Galactic source as, in case of an 
AGN, the optical would be also completely absorbed, while a Galactic stellar component 
could have significant emission in optical and not in the UV domain. 

\begin{table}[htbp]
\caption{Summary of the possible type for each counterpart of the seventeen sources, obtained 
through the analysis presented in this paper.}
\label{tab:results}
\begin{tabular}{ll} 
\hline\hline             
Name &   Type \& Comment\\
(IGR) &     \\ 
\hline
J03184$-$0014 & IGR and Swift sources not related\\
J03532$-$6829 &  $z$=0.087 BL Lac\\
J05319$-$6601 & probable XRB in LMC  \\
J05346$-$5759 & CV (not an IP?)  \\
J09025$-$6814 & AGN, poss. Compton thick, Sey 2(?) \\
J10101$-$5654 & HMXB  \\
J13000+2529   & AGN \\
J13020$-$6359 & HMXB with pulsar\\
J15161$-$3827 \#1& AGN, Liner/Sey 2 \\
               \#2    & ? \\
 \#3 & YSO \\
 \#4 & ?\\
J15479$-$4529 &   CV/IP\\
J16287$-$5021 &   HMXB (?)\\
J17353$-$3539 &   HMXB (?)\\
J17476$-$2253 &  XRB (?) \\
J18214$-$1318 &  probable HMXB (sg star?) \\
J19267+1325   & Galactic source \\
J20286+2544 \#1  & AGN, Sey 2\\
            \#2    & AGN, Sey 2 (?)\\
J23206+6431   &  AGN, Sey 1 \\
\hline
\hline
\end{tabular}
\end{table}

\section{Summary and conclusions}
In this paper, we reported the X-ray analysis of seventeen hard X-ray sources discovered 
by \integral. The refined X-ray positions provided by the \swift\ observations 
(Table~\ref{tab:position}) allowed us to pinpoint the possible IR and optical counterparts 
in most of the cases. Table~\ref{tab:results} reports the conclusions of our analysis 
concerning the possible type of each of the seventeen sources. We confirm the associations 
and types previously suggested for five sources: \\
\indent$\bullet$ IGR J03532$-$6829 is a BL Lac\\
\indent$\bullet$ IGR J05346$-$5759 and J15479$-$4529 are  CVs, the latter is an IP\\
\indent$\bullet$ IGR J10101$-$5654   is very likely an  HMXB\\ 
\indent$\bullet$ IGR J18214$-$1318 is a  probable HMXB \\
\indent$\bullet$ IGR J13000+2529 and J23206+6431 are AGNs. The latter is a Sey~1 \\
\indent$\bullet$ IGR J13020$-$6359 is an HMXB containing a pulsar\\
\indent In 2 cases, we detected several X-ray counterparts in the IBIS error circle. In 
these cases, the spectral analysis of each of those sources allowed us to suggest that 
Swift J151559.3$-$382548 is a probable Sey 2 AGN, which is the likely counterpart 
to IGR J15161$-$3827. In the case of IGR~J20286+2544, the \swift\ error circle contains 
two AGNs, and the \integral\ source seems to be a blend of those two objects, although 
Swift J202834.9+254359 (=MCG+04-48-002) is brighter and therefore contributes more to 
the hard X-ray emission.\\
\indent In one case (IGR J19267+1325),  we do not detect any X-ray source within the IBIS 
error circle. A bright source, however, has a position that is marginally consistent, and,
although it is slightly outside the IBIS error circle, our analysis leads us to suggest 
that both sources are related. We could not unambiguously unveil its true nature, although we 
favoured a Galactic source.\\
\indent Of the six remaining source:\\
\indent$\bullet$ IGR J05319$-$6601 is compatible with being an X-ray binary in the LMC\\
\indent$\bullet$ We identified IGR J09025$-$6814 with the nucleus of a galaxy, and 
provided the first identification of this source as an AGN and a possible Sey 2\\
\indent$\bullet$ We suggest that IGR~J16287$-$5021, J17353$-$3539 and J17476$-$2253 are
probable X-ray binaries and possibly HMXBs.\\
\indent$\bullet$ We find an X-ray source slightly outside the IBIS error circle 
of IGR~J03184$-$0014, but our analysis does not favour any association between
the \swift\ and \integral\ objects.\\

\begin{acknowledgements}
JR thanks the Swift help desk for their great help and rapid answer.
JAT acknowledges partial support from a NASA INTEGRAL Guest Observer grant 
NNX07AQ13G. We warmly thank the anonymous referee for his/her very constructive
comments, that really helped to improve to quality of this paper.
We acknowledge the use of data collected with the \swift\ observatory.
This research has made use of the USNOFS Image and Catalogue Archive
operated by the United States Naval Observatory, Flagstaff Station
(http://www.nofs.navy.mil/data/fchpix/)
 This research has made use of the SIMBAD database, operated at CDS, Strasbourg, France.
It also makes use of data products from the Two Micron All Sky Survey, which 
is a joint project of the University of Massachusetts and the Infrared Processing 
and Analysis Center/California Institute of Technology, funded by the National 
Aeronautics and Space Administration and the National Science Foundation.
This research has made use of the NASA/IPAC Extragalactic Database (NED) which is 
operated by the Jet Propulsion Laboratory, California Institute of Technology, under 
contract with the National Aeronautics and Space Administration. 

\end{acknowledgements}

\bibliography{ms}
\end{document}